\documentclass[12pt]{article}
\usepackage{hyperref}
\usepackage{url}
\usepackage{listings,jvlisting}
\usepackage{amssymb}
\usepackage{tabularx}
\usepackage{amsmath}
\usepackage{mdframed}
\usepackage{graphicx}
\usepackage{setspace}
\usepackage{here}
\usepackage{authblk}
\usepackage{mathtools}
\usepackage{amsthm}
\usepackage{algorithmic}
\usepackage{algorithm}
\usepackage{cite}
\usepackage{comment}

\def\al{{\alpha}}
\def\be{{\beta}}
\def\ga{{\gamma}}
\def\de{{\delta}}
\def\ep{{\varepsilon}}
\def\la{{\lambda}}

\def\si{{\sigma}}

\def\alh{{\widehat \al}}

\def\beh{{\hat \be}}

\def\nh{{\hat n}}

\def\q{{\text{\boldmath $q$}}}

\def\ph{{\hat p}}
\def\qh{{\hat q}}

\def\dt{{\tilde d}}

\def\Nc{{\cal N}}

\begin{document}
\title{Hybrid Non-informative and Informative Prior Model-assisted Designs for Mid-trial Dose Insertion}
%著者順は今後相談，
%最高用量より上の用量を追加することについても触れるべきかどうか？どの程度触れるべきか？
\author[1]{Kana Yamada}
\author[2]{Hisato Sunami}
\author[4]{Kentaro Takeda}
\author[5]{Keisuke Hanada}
\author[3,6]{Masahiro Kojima\footnote{Address:1-13-27 Kasuga,Bunkyo-ku,Tokyo 112-8551, Japan. Tel: +81-(0)3-3817-1949 \quad
E-Mail: mkojima263@g.chuo-u.ac.jp}}
\affil[1]{Institute of Science Tokyo}
\affil[2]{The University of Osaka}
\affil[3]{Chuo University}
\affil[4]{Astellas Pharma Global Development Inc.}
\affil[5]{Wakayama Medical University}
\affil[6]{The Institute of Statistical Mathematics}

\maketitle
\abstract{\noindent
In oncology phase I trials, model-assisted designs have been increasingly adopted because they enable adaptive yet operationally simple dose adjustment based on accumulating safety data, leading to a paradigm shift in dose-escalation methodology. In practice, a single mid-trial dose insertion may be considered to examine safer doses and/or to collect more informative efficacy data. In this study, we investigate methods to improve dose assignment and the selection of the maximum tolerated dose (MTD) or the optimal biological dose (OBD) when a new dose level is added during an ongoing trial under a model-assisted framework, by assigning informative prior information to the inserted dose. We propose a hybrid design that uses a non-informative model-assisted design at trial initiation and, upon dose insertion, applies an informative-prior extension only to the newly added dose. In addition, to address potential skeleton misspecification, we propose two adaptive extensions: (i) an online-weighting approach that updates the skeleton over time, and (ii) a Bayesian-mixture approach that robustly combines multiple candidate skeletons. We evaluate the proposed methods through simulation studies.
}
\par\vspace{4mm}

\section{Introduction}
\begin{figure}[htbp]
  \centering
  \includegraphics[width=0.6\linewidth]{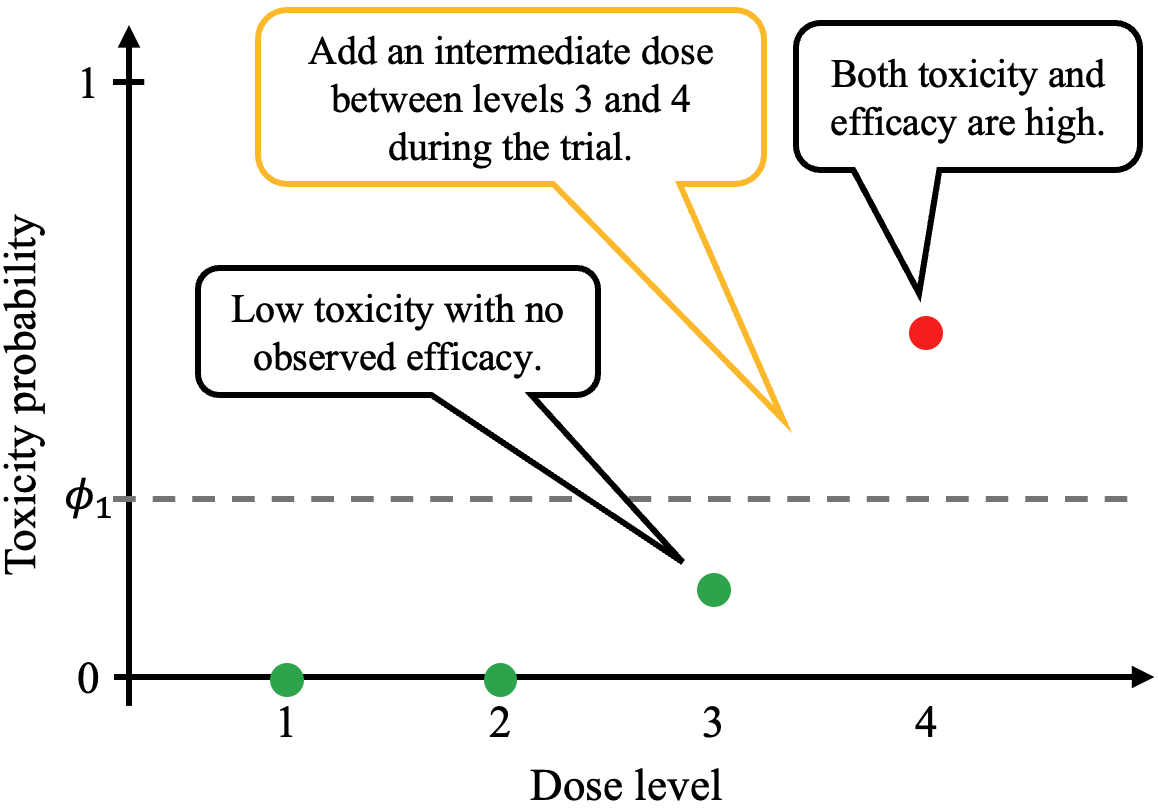}
  \caption{An Example of the Study Motivation}
  \label{fig:Example_figure}
\end{figure}

In recent years, model-assisted designs such as the Bayesian Optimal Interval (BOIN) design~\cite{liu2015boin} have been increasingly adopted in oncology phase I trials. These designs can select the maximum tolerated dose (MTD) accurately while remaining operationally simple. Dose levels in oncology phase I studies are often selected based on nonclinical data when there is no prior human dosing experience, as in first-in-human trials. However, if the initial dose selection based on nonclinical data proves inadequate, it may become necessary to insert an additional dose level between adjacent dose levels during the ongoing trial. To illustrate the concept with a simple example, as shown in Figure \ref{fig:Example_figure}, let $\phi_1$ denote the target toxicity level (TTL). 
If no efficacy is observed at dose level 3 and its toxicity probability is below $\phi_1$, whereas at dose level 4 the toxicity probability exceeds $\phi_1$ with efficacy observed, it is natural to consider evaluating an additional dose level between dose levels 3 and 4. Next, we present three examples of such mid-trial dose insertion from actual clinical trials.

\begin{enumerate}
    \item In a first-in-human Phase I trial of futibatinib (TAS-120), an intermediate 20 mg QD cohort was added per a protocol amendment dated 15 May 2017~\cite{bahleda2020phase}. According to the protocol and the published report, the Phase I QD dose-escalation evaluated 4, 8, 16, 20, and 24 mg; 3/9 evaluable patients had DLTs at 24 mg during cycle 1, whereas 0/5 evaluable patients had DLTs at 20 mg. Consequently, 20 mg QD was determined as the maximum tolerated dose and supported as the recommended Phase II dose in this program.
    \item An open-label phase~1/2 dose-escalation study of intratumourally administered INT230-6 initially planned six monotherapy cohorts---five intrapatient dose-escalation cohorts (A1, B1, EA, EC, EC2) and a subsequently added fixed-dose cohort (EC3). In terms of maximum total injection volume, the EC2 cohort allowed up to 220~mL and was numerically the highest, whereas the subsequently added EC3 cohort used a maximum total volume of 175~mL per treatment session. Although the study did not reach an MTD, the investigators added EC3 (sixth dose level; $n=9$) to collect additional safety and preliminary efficacy information~\cite{thomas2025safety}. From the publication, it is not entirely clear which dose regimen was ultimately selected for further development.
    \item The SAP for the PF-04449913 (glasdegib) study B1371005 (NCT02038777) prespecified the originally planned dose levels and also allowed dose insertion when clinically and scientifically justified---specifically, the insertion of an intermediate dose (e.g., 80~mg once daily) and the addition of a dose below the planned minimum (e.g., 25~mg once daily). In practice, 25~mg once daily was added with the aim of assessing pharmacodynamic effects, whereas 80~mg once daily was not added in B1371005~\cite{pfizer2019b1371005sap,minami2017phase}.
\end{enumerate}

Regarding statistical considerations for dose insertion, Hu et al. (2013)~\cite{hu2013adaptive} proposed a method to add a new dose level when the prespecified doses are estimated to fall outside an acceptable toxicity region. Guo et al. (2015)~\cite{guo2015teams} proposed a phase I/II design, referred to as the Toxicity and Efficacy Based Dose Insertion Design (TEAMS), with the objective of selecting the optimal biological dose (OBD). In TEAMS, dose insertion is triggered when the credible interval for the estimated optimal dose does not contain any of the existing dose levels, and the phase I component is implemented based on the mTPI design. However, examples from actual clinical trials indicate that newly added dose levels may be selected in an ad hoc manner based on clinical and nonclinical considerations.  More recently, George et al. (2026)~\cite{george2026novel} proposed a general framework for mid-trial dose insertion in early-phase oncology combination trials, where posterior inference on the target-toxicity contour is used to assess whether any prespecified dose combination is close to the TTL; if not, new dose combinations are inserted and the model is updated accordingly. However, none of Hu et al. (2013)~\cite{hu2013adaptive}, Guo et al. (2015)~\cite{guo2015teams}, or George et al. (2026)~\cite{george2026novel} provides a principled approach to constructing prior information for an inserted dose by borrowing information from adjacent doses. In the setting considered in this study, namely insertion of an intermediate dose, substantial experience has already accumulated at the adjacent doses, and toxicity has also been observed at doses higher than the inserted dose. Therefore, it is important to leverage information from neighboring doses to construct prior information for the newly inserted intermediate dose.

In this study, we investigate methods to improve the operating characteristics for selecting the MTD when new dose levels are added during an ongoing trial, by borrowing toxicity information from the adjacent dose levels administered before and after the newly inserted dose. In addition, motivated by Project Optimus~\cite{fda2023optimus} and the recent shift in focus from the MTD alone to the selection of an optimal biological dose (OBD) based on utility that jointly considers toxicity and efficacy, we also describe how the proposed approach can be extended to trials conducted under the Bayesian optimal interval design for dose finding based on both efficacy and toxicity outcomes (BOIN-ET) design~\cite{takeda2018boin}.

This paper is organized as follows. Section 2 presents the preliminaries. Section 3 describes the proposed methods. Section 4 presents the simulation settings and results. Section 5 concludes with a discussion.

\section{Preliminaries}
In a Phase I oncology clinical trial, we assume a prespecified set of $J$ dose levels planned at the start of the trial. These dose levels are determined based on nonclinical studies. The total planned sample size is denoted by $N$. When selection of the MTD is one of the primary objectives, the TTL is denoted by $\phi_1$. In this study, we consider applying a model-assisted design to a dose-escalation trial and assume that dose assignment is adaptively modified based on accumulating toxicity outcomes. As illustrated in Figure \ref{fig:Example_figure}, when a dose level is judged to require de-escalation because of excessive toxicity despite showing evidence of efficacy, while the next lower dose level exhibits neither toxicity nor efficacy, we consider adding an intermediate dose during the trial.

A key feature of our setting is that, at the time of dose insertion, treatment experience has already been obtained at the existing dose levels, providing a small amount of information on toxicity and efficacy. Therefore, for the newly added dose, we propose a hybrid design that applies an informative model-assisted design incorporating information from adjacent dose levels, while the remaining dose levels continue to be governed by a non-informative model-assisted design.

We begin with a brief review of representative non-informative model-assisted designs, including BOIN and BOIN-ET. We then provide an overview of their informative extensions, namely the Bayesian Optimal Interval Design with Informative Prior (iBOIN)~\cite{zhou2021incorporating} and Incorporating historical information to improve dose optimization design with toxicity and efficacy endpoints (iBOIN-ET) designs~\cite{zhao2023incorporating}.

The following notation is used throughout all designs. Let $n_j$ denote the number of patients treated at dose level $j$, and $t_j$ the number of patients who experience a dose-limiting toxicity (DLT) at that dose level.
The true DLT probability at dose level $j$ is denoted by $p_j$, and its estimate is $\hat p_j = t_j / n_j$.

\medskip
\noindent
{\bf [BOIN]}\\
The BOIN design defines three hypotheses for the true toxicity probability $p_j$, where $\phi_2 < \phi_1 < \phi_3$: equal to the TTL ($H_1: p_j=\phi_1$), less than the TTL ($H_2: p_j=\phi_2$), and greater than the TTL ($H_3: p_j=\phi_3$). It is an optimal design that minimizes the probability of making an incorrect dose adjustment under each hypothesis. As the optimal solution, decision boundaries $(\la_e,\la_d)$ are derived, and dose adjustment is performed by comparing the observed toxicity probability $\ph_j$ with these boundaries. 
\begin{align}
\la_e=\frac{\log{\left(\frac{1-\phi_2}{1-\phi_1}\right)}}{\log\left\{\frac{\phi_1(1-\phi_2)}{\phi_2(1-\phi_1)}\right\}},
\la_d=\frac{\log{\left(\frac{1-\phi_1}{1-\phi_3}\right)}}{\log\left\{\frac{\phi_3(1-\phi_1)}{\phi_1(1-\phi_3)}\right\}},\nonumber
\end{align}
Specifically, if $\ph_j \in (\la_e,\la_d)$, the next dose level remains $j$; if $\ph_j \leq \la_e$, the next dose level is $j+1$; and if $\ph_j \geq \la_d$, the next dose level is $j-1$.

\medskip
\noindent
{\bf [BOIN-ET]}\\
The BOIN-ET design is an extension of the BOIN design that incorporates an efficacy threshold, allowing dose assignment to be guided by both safety and efficacy. Let the target efficacy probability threshold be denoted by $\eta$. For dose level $j$, let $u_j$ denote the number of efficacy responses, $q_j$ the true efficacy probability, and $\qh_j = u_j / n_j$ the observed efficacy probability. The dose assignment rules are as follows: if $\ph_j \leq \la_1$ and $\qh_j \leq \eta$, the next cohort is treated at dose $j+1$; if $\ph_j \leq \la_2$ and $\qh_j > \eta$, the next cohort remains at dose $j$; if $\ph_j \geq \la_2$, the next cohort is de-escalated to dose $j-1$. In cases not covered by these rules, the next cohort is assigned to the dose with the highest efficacy probability among $(j-1, j, j+1)$. If two or more doses share the highest efficacy probability, one of them is randomly selected. The dose-escalation and de-escalation boundaries $(\la_1,\la_2,\eta)$ are obtained by minimizing the probability of incorrect dose-assignment decisions under the following six hypotheses.
\begin{align}
&H_{11j}:p_j=\phi_1\mbox{ and }q_j=\de_1, H_{12j}:p_j=\phi_1\mbox{ and }q_j=\de_2,\nonumber\\
&H_{21j}:p_j=\phi_2\mbox{ and }q_j=\de_1, H_{22j}:p_j=\phi_2\mbox{ and }q_j=\de_2,\nonumber\\
&H_{31j}:p_j=\phi_3\mbox{ and }q_j=\de_1, H_{32j}:p_j=\phi_3\mbox{ and }q_j=\de_2,\nonumber
\end{align}
where $\de_1$ denotes the probability corresponding to the target efficacy, and $\de_2$ indicates that the efficacy probability is lower than the target probability.

\medskip
In this study, we consider a situation in which efficacy is observed at a certain dose level but dose de-escalation is required due to excessive toxicity, whereas at the next lower dose level, no toxicity is observed but no efficacy is observed either. Under this situation, we consider inserting an additional dose level between these two doses. Although the dose-adjustment designs introduced above are non-informative, in the setting considered in this study, there already exist dose levels for which treatment experience has been accumulated. Accordingly, we propose a hybrid design that leverages information obtained from previously treated dose levels by applying an informative prior to the newly added dose, while continuing to use non-informative priors for the existing dose levels. We first introduce designs that apply informative priors. To borrow information, it is necessary to specify the skeleton, which represents the prior distribution of toxicity probabilities, and the prior effective sample size (PESS), which reflects the sample size of the prior distribution. Let the skeleton of the newly inserted dose be denoted by $r$, and the PESS by $s$. The methods for determining $r$ and $s$ will be proposed later.

\medskip
\noindent
{\bf [iBOIN]}\\
We consider applying informative prior distributions to the probabilities associated with each of the three hypotheses introduced in the BOIN design. Let $\pi_1$ denote the probability that hypothesis $H_1$ holds, $\pi_2$ the probability that hypothesis $H_2$ holds, and $\pi_3$ the probability that hypothesis $H_3$ holds. For $k = 1, 2, 3$, the informative prior distribution $\pi^\ast_k$ is given as follows.
\begin{align}
\pi^\ast_k=\sum^{s}_{x=0}w_1(x,\phi_k)\binom{s}{x}(r)^x(1-r)^{s-x},\nonumber
\end{align}
where $w_1(x,\phi_k)=\frac{\phi_k^x(1-\phi_k)^{s-x}}{\sum^3_{k^\prime=1}\phi_{k^\prime}^x(1-\phi_{k^\prime})^{s-x}}$. The dose-escalation and de-escalation boundaries are given as follows (in the BOIN design, $\pi_1 = \pi_2 = \pi_3 = 1/3$ was assumed). Note that in the following, we present the boundaries for a dose that is added during the trial.
\begin{align}
\la^\ast_e=\frac{\log{\left(\frac{1-\phi_2}{1-\phi_1}\right)}+\frac{1}{n}\log{\left(\frac{\pi^\ast_2}{\pi^\ast_1}\right)}}{\log\left\{\frac{\phi_1(1-\phi_2)}{\phi_2(1-\phi_1)}\right\}},
\la^\ast_d=\frac{\log{\left(\frac{1-\phi_1}{1-\phi_3}\right)}+\frac{1}{n}\log{\left(\frac{\pi^\ast_1}{\pi^\ast_3}\right)}}{\log\left\{\frac{\phi_3(1-\phi_1)}{\phi_1(1-\phi_3)}\right\}},\nonumber
\end{align}
where $n$ denotes the actual number of patients treated at the dose that was added during the trial. For the dose to which iBOIN is applied, dose adjustment is conducted using $\lambda_e^\ast$ and $\lambda_d^\ast$. In addition, to incorporate prior information when selecting the MTD, we use $\ph=\frac{t+sr}{n+s}$, where $t$ denotes the number of observed DLTs at the added dose. The PESS is recommended to select $\frac{N}{3J}$ or $\frac{N}{2J}~$\cite{zhou2021incorporating}.

\medskip
\noindent
{\bf [iBOIN-ET]}\\
In the iBOIN-ET design~\cite{zhao2023incorporating}, it is necessary to specify not only the skeleton for toxicity probabilities but also the skeleton for efficacy probabilities, denoted by $v$. As in the BOIN design, prior information is incorporated under the six hypotheses. The prior information can be calculated as follows. 
\begin{align}
\pi_{km}=\sum^{s}_{x,y=0}w_1(x,\phi_k)w_2(y,\de_m)\binom{s}{x}(r)^x(1-r)^{s-x}\binom{s}{y}(v)^y(1-v)^{s-y},\nonumber
\end{align}
where $w_2(y,\de_m)=\frac{\de_m^y(1-\de_m)^{s-y}}{\sum^2_{m^\prime=1}\de_{m^\prime}^y(1-\de_{m^\prime})^{s-y}}$. In this paper, we adopt the M-iBOIN-ET design~\cite{zhao2023incorporating}, in which the thresholds can be calculated more easily, although the proposed method can also be applied to the J-iBOIN-ET design. The three thresholds are given as follows.
\begin{align}
&\la^\ast_{e,ET}=\frac{\log{\left(\frac{1-\phi_2}{1-\phi_1}\right)}+\frac{1}{n}\log{\left(\frac{\pi^\ast_{21}+\pi^\ast_{22}}{\pi^\ast_{11}+\pi^\ast_{12}}\right)}}{\log\left\{\frac{\phi_1(1-\phi_2)}{\phi_2(1-\phi_1)}\right\}}, \la^\ast_{d,ET}=\frac{\log{\left(\frac{1-\phi_1}{1-\phi_3}\right)}+\frac{1}{n}\log{\left(\frac{\pi^\ast_{11}+\pi^\ast_{12}}{\pi^\ast_{31}+\pi^\ast_{32}}\right)}}{\log\left\{\frac{\phi_3(1-\phi_1)}{\phi_1(1-\phi_3)}\right\}},\nonumber\\
&\eta^\ast=\frac{\log{\left(\frac{1-\de_2}{1-\de_1}\right)}+\frac{1}{n}\log{\left(\frac{\pi^\ast_{12}+\pi^\ast_{22}+\pi^\ast_{32}}{\pi^\ast_{11}+\pi^\ast_{21}+\pi^\ast_{31}}\right)}}{\log\left\{\frac{\de_1(1-\de_2)}{\de_2(1-\de_1)}\right\}}.\nonumber
\end{align}
The recommended PESS is $\frac{N}{2J}$ or $\frac{N}{J}$~\cite{zhao2023incorporating}.

\section{Proposed method}
In this study, we propose methods for determining the toxicity and efficacy skeletons of a dose $d^\ast$ added during the course of the trial based on the toxicity and efficacy information from the existing dose levels. 
In addition, we propose a method for determining the effective sample size (ESS) of the posterior distribution, rather than the PESS, which takes into account the numbers of patients treated at other dose levels; accordingly, we refer to this quantity as ESS.

We first introduce the method for calculating the toxicity skeleton. 
To reflect all available information from the existing dose levels, the toxicity skeleton is estimated by fitting Bayesian Logistic Regression Model (BLRM) for the toxicity probability $p_T$.
\begin{align}
\label{eq:BLRM}
\mbox{logit}(p_T(d_j))=\al+\be\log\left(\frac{d_j}{d_{ref}}\right)
\end{align}
The skeleton for the newly added dose $d^\ast$ is provided by the model estimated using all the data available just before the addition of $d^\ast$ as $(1+\exp({-\alh-\beh\log\left(d^\ast/d_{ref}\right)}))^{-1}$, where $\alh$ and $\beh$ are the posterior mean, $d_{ref}$ is chosen as a clinically relevant dose at which the MTD was anticipated, and was used as a scaling constant to facilitate prior specification and numerical stability. Because the BLRM is applied only when determining the dose skeleton for newly added dose levels, the transparency of the BOIN and BOIN-ET designs is not compromised, and the computational and analytical burden remains lower than that of fully model-based designs.

For the efficacy skeleton, since the probability of efficacy may not be necessarily monotone increasing with dose, we consider by fitting a second-degree fractional polynomial logistic regression model~\cite{royston1994regression} for the efficacy probability $p_E$,
\begin{align}
\mathrm{logit}\{p_E(d_j)\}
=\al+\be_1 f_1\!\left(d_j^\dagger, k_1\right)+\be_2 f_2\!\left(d_j^\dagger, k_1, k_2\right),\nonumber
\end{align}
where $d_j^\dagger = d_j/\bar d$ is the standardized dose, $\bar d$ denotes the average of all planned dose levels,
and $k_1,k_2 \in \{-2,-1,-0.5,0,0.5,1,2\}$ are power parameters. The basis functions $f_1(\cdot)$ and $f_2(\cdot)$ are defined as
\begin{align}
f_1\left(d_j^\dagger, k_1\right)
=
\begin{cases}
\log\left(d_j^\dagger\right), & k_1 = 0, \\
\left(d_j^\dagger\right)^{k_1}, & k_1 \neq 0,
\end{cases}\nonumber
\end{align}
and
\begin{align}
f_2\left(d_j^\dagger, k_1, k_2\right)
=
\begin{cases}
\left\{\log\left(d_j^\dagger\right)\right\}^2, 
& k_1 = k_2 = 0, \\
\left(d_j^\dagger\right)^{k_1}\log\left(d_j^\dagger\right), 
& k_1 = k_2 \neq 0, \\
\log\left(d_j^\dagger\right), 
& k_2 = 0,\ k_1 \neq k_2, \\
\left(d_j^\dagger\right)^{k_2}, 
& k_2 \neq 0,\ k_1 \neq k_2.
\end{cases}\nonumber
\end{align}
To select the optimal power parameters $(\hat k_1, \hat k_2)$, we use the binomial deviance rather than the Akaike Information Criterion, because the penalty term is constant across candidate models and therefore does not affect model comparison. Specifically, for each candidate pair $(k_1,k_2)$, the model parameters $(\alpha,\beta_1,\beta_2)$ are estimated by maximum likelihood, and the deviance is defined as
\begin{align}
D(k_1,k_2)
=-2\left\{\ell(\hat\alpha,\hat\beta_1,\hat\beta_2 \mid k_1,k_2)-\ell_{\mathrm{sat}}\right\},\nonumber
\end{align}
where $\ell(\hat\alpha,\hat\beta_1,\hat\beta_2 \mid k_1,k_2)$ denotes the maximized binomial log-likelihood under the fractional polynomial model,
and $\ell_{\mathrm{sat}}$ is the saturated log-likelihood corresponding to a model that assigns an unconstrained success probability to each observed dose level. The saturated log-likelihood does not depend on $(k_1,k_2)$. The optimal power parameters $(\hat k_1,\hat k_2)$ are chosen as the pair minimizing
$D(k_1,k_2)$, and the resulting model is used to derive the efficacy skeleton.

Next, since the outcome is binary, an ESS is required; therefore, we introduce a method for calculating the ESS.  Unlike the PESS used in the iBOIN and iBOIN-ET designs, we consider a data-dependent ESS in this study. Accordingly, our approach is based on the approximation formula using the first and second moments proposed by Zhou et al. (2021)~\cite{zhou2021incorporating}. We first present the ESS for toxicity, which is given as follows. Specifically, after fitting the BLRM in (\ref{eq:BLRM}) using all data available immediately before the insertion of the new dose $d^\ast$, we obtain posterior samples of the toxicity probability
$p_T(d^\ast)$.
Let
\[
\mu^\ast_T = \mathbb{E}\{p_T(d^\ast)\mid \mathcal{D}\},
\qquad
v^\ast_T = \mathrm{Var}\{p_T(d^\ast)\mid \mathcal{D}\},
\]
denote the posterior mean and variance, respectively, where $\mathcal{D}$ represents the accumulated data prior to dose insertion.

To translate the posterior uncertainty into an effective prior sample size, we approximate the posterior distribution of $p_T(d^\ast)$ by a Beta distribution.
For a Beta$(a,b)$ distribution, the mean and variance are given by $\mu = \frac{a}{a+b}$ and $v = \frac{\mu(1-\mu)}{a+b+1}$. Solving for $a+b$ yields $a+b = \frac{\mu(1-\mu)}{v} - 1$. Accordingly, we define the ESS of toxicity for the inserted dose as
\begin{align}
n^{\ast}_{T}=\frac{\mu_T^\ast(1-\mu_T^\ast)}{v_T^\ast} - 1.\nonumber
\end{align}

To avoid excessive borrowing and to improve robustness against potential model misspecification, following the approach adopted in the iBOIN and iBOIN-ET designs, we introduce a scaling factor $\ga_1\in(0,1]$ and define the adjusted ESS as
\begin{align}
\nh^\ast_T = \ga_1 \, n^{\ast}_{T}.\nonumber
\end{align}
Finally, to ensure numerical stability and to prevent the prior information from overwhelming the observed data, the ESS is truncated to the range $0 \le \nh^\ast_T \le \max_j n_j$, where $n_j$ denotes the number of patients treated at dose level $j$ prior to dose insertion.

For efficacy, we calculate the ESS in the same way as for toxicity.
Specifically, after fitting the efficacy model used to construct the efficacy skeleton
to all data available immediately before the insertion of the new dose $d^*$,
we obtain posterior samples of the efficacy probability $p_E(d^*)$.
Let
\begin{align}
\mu_E^* = \mathbb{E}\{ p_E(d^*) \mid \mathcal{D} \}, \qquad
v_E^* = \mathrm{Var}\{ p_E(d^*) \mid \mathcal{D} \},\nonumber
\end{align}
denote the posterior mean and variance, respectively, where $\mathcal{D}$ represents
the accumulated data prior to dose insertion.

As in the toxicity case, we approximate the posterior distribution of $p_E(d^*)$
by a Beta distribution.
Using the relationship between the mean and variance of a Beta distribution,
the ESS for efficacy is defined as
\begin{align}
n_E^* = \frac{\mu_E^*(1-\mu_E^*)}{v_E^*} - 1.\nonumber
\end{align}
The same scaling factor $\ga_2\in(0,1]$ and truncation rule as those used for toxicity
are applied to $n_E^*$.

\subsection{Data-Driven Updating of the Skeleton and ESS}
As treatment experience accumulates at a newly added dose level, it may become apparent that the initially specified skeleton differs materially from the underlying true toxicity profile. In Bayesian inference, even when the prior distribution is not well aligned with the data, the impact of the prior typically decreases as more data accrue. However, in the setting considered in this study, an informative skeleton is applied only to the single newly inserted dose, and Phase I oncology trials usually treat only a small number of patients at any given dose level (often on the order of 9–12). Under such limited sample sizes, discrepancies between the skeleton and the observed data may remain influential, leading to biased inference and potentially misleading dose recommendations.

To address this issue, we propose two approaches based on an online learning method and a Bayesian mixture model. The first approach, referred to as the online updating–based approach, recalculates the skeleton using data obtained after dose insertion and adaptively learns whether the previously specified skeleton or the updated skeleton better explains the observed data. The second approach adopts a Bayesian mixture approach that combines multiple candidate skeletons, thereby improving robustness to skeleton misspecification. Below, we describe the procedure for updating the toxicity skeleton. Although online learning and Bayesian mixture are used to update the model, such updates are required only when dose-adjustment decisions are made for the newly inserted dose level. As a result, the proposed approach imposes substantially less computational and operational burden than fully model-based designs that require continuous updating. Moreover, because the framework is built upon iBOIN and iBOIN-ET, the simplicity and transparency inherent to model-assisted designs are preserved.
The efficacy skeleton can be updated in an analogous manner, and thus the detailed formulas are omitted.

\medskip
\noindent
[{\bf Online learning approach}]\\
We formulate the selection of the skeleton for the inserted dose as an online learning problem with multiple candidate models. At each update time $t$, we consider two candidate skeletons for the inserted dose $d^\ast$: (i) the previously specified skeleton $r^{\ast}_{0,t}$, which remains unchanged from the last update, and (ii) an updated skeleton $r^{\ast}_{1,t}$ obtained from the current BLRM fit. Let $y^\ast_t$ denote the number of DLTs observed among $n^\ast_t$ patients treated at dose $d^\ast$ during the interval between updates.
For a candidate toxicity probability $p_T \in (0,1)$, we use the binomial logarithmic loss (negative log-likelihood),
\begin{align}
\ell_t(p_T)&=-\log \Pr\!\left(Y^\ast_t = y^\ast_t \mid n^\ast_t, p_T\right) \nonumber\\
&=- y^\ast_t \log p_T - (n^\ast_t - y^\ast_t)\log(1 - p_T) + C_t,\nonumber
\end{align}
where $C_t$ is a constant that does not depend on $p_T$. The exponentially weighted forecaster, also known as the Hedge algorithm~\cite{freund1997decision}, updates the weights assigned to the two
candidate skeletons according to
\begin{align}
w_{k,t+1}
=\frac{w_{k,t} \exp\!\left\{-\ell_t\!\left(r^{\ast}_{k,t}\right)\right\}}
{\sum_{j \in \{0,1\}} w_{j,t} \exp\!\left\{-\ell_t\!\left(r^{\ast}_{j,t}\right)\right\}},
\qquad k \in \{0,1\}.\nonumber
\end{align}
The initial weights $w_{0,1}$ and $w_{1,1}$ are assumed to be positive and satisfy $w_{0,1} + w_{1,1} = 1$. The skeleton used for information borrowing at the next decision point is then defined as the convex combination
\begin{align}
r^\ast_t=w_{0,t}\, r^{\ast}_{0,t}+w_{1,t}\, r^{\ast}_{1,t} .\nonumber
\end{align}
This procedure corresponds to a standard online learning strategy for probabilistic prediction under logarithmic loss
\cite{cesa-bianchi2006prediction, shalev2011online}.
Moreover, under logarithmic loss, the exponential-weight update coincides with Bayesian model averaging (Bayes mixture)
when each candidate skeleton is interpreted as a probabilistic model and the initial weights $w_{k,1}$ are viewed as
prior model weights \cite{cesa-bianchi2006prediction}.

Alternatively, one may consider selecting a single skeleton from the candidate skeletons.
In this case, at each update time, a single skeleton can be selected by choosing the candidate with the smaller cumulative logarithmic loss (Follow-the-Leader, FTL):
\begin{align}
\hat k_{t}
=\arg\min_{k\in\{0,1\}}\sum_{s=1}^{t-1} \ell_s\!\left(r^{\ast}_{k,s}\right),
\qquad
r^\ast_t = r^{\ast}_{\hat k_t,t}.\nonumber
\end{align}
FTL (and closely related best-expert selection rules) is a classical online learning approach
\cite{cesa-bianchi2006prediction, shalev2011online}.
Because hard switching can be unstable under small sample sizes, soft update strategies may be preferred for improved stability.
Switching penalties or tracking variants (e.g., fixed-share or tracking the best expert) can also be introduced to limit frequent changes \cite{herbster1998tracking}.

\medskip
\noindent
[{\bf Bayesian mixture approach}]\\
In the updating-based approach, we compare the updated skeleton for the inserted dose with the previously specified skeleton to determine which better explains the observed data.
In contrast, the mixture approach considers multiple candidate skeletons—including the skeleton for the inserted dose and those derived from adjacent dose levels—and evaluates their relative support from the observed data. The first skeleton ($r_1$) is derived from equation (\ref{eq:BLRM}).
The second skeleton ($r_2$) represents the toxicity probability at the dose level immediately above the newly added dose, and the third skeleton ($r_3$) represents the toxicity probability at the dose level immediately below it.
The mixture prior ($r_w$) is defined as
\begin{align}
r_w = w_1 r_1 + w_2 r_2 + w_3 r_3, \quad w_1 + w_2 + w_3 = 1.\nonumber
\end{align}
We place a Dirichlet prior on the mixing weights $(w_1, w_2, w_3)$. Under a noninformative setting, we use a uniform Dirichlet$(1,1,1)$ prior. After administering the new dose $d^\ast$, let $n^\ast$ and $y^\ast$ denote the number of treated patients and the number of observed toxicities, respectively. For each candidate skeleton, the predictive likelihood is defined as $m_k = p(y^\ast \mid r_k)$, and the posterior weights are computed as
\begin{align}
w_k^{post} = \frac{w_k^{prior} m_k}{\sum_{l=1}^3 w_l^{prior} m_l}, \qquad k\in\{1,2,3\}.\nonumber
\end{align}
We take $m_k=\Pr(Y^\ast=y^\ast\mid n^\ast,r_k)$ under a Binomial$(n^\ast,r_k)$ model.
A single skeleton may be selected by choosing the candidate corresponding to the largest posterior weight.
Alternatively, the mixture prior $r_w$ obtained by replacing $w_k$ with $w_k^{post}$ may be used.

\medskip
For each approach, because the ESS of toxicity reflects the amount of information accumulated in the data, it can be updated at each update time by recalculating it using all currently available data via the same procedure described above. To ensure that the amount of borrowed information does not exceed that of the empirical data actually informing the skeleton, the ESS is capped at the largest sample size among the dose levels used in the skeleton estimation.

\section{Simulation}
%初期用量で一例目からスタートするわけではなく，ある程度投与が終えている特定の条件下からスタートするシミュレーションにする？
\subsection{Simulation configuration}
To investigate the performance of the proposed hybrid approaches of BOIN/iBOIN and BOIN-ET/iBOIN-ET, the true MTD selection rate and/or true OBD selection rate were compared with conventional BOIN and BOIN-ET. We considered a phase I dose-finding setting with five dose levels
$d_1,\dots,d_5$ corresponding to 300, 900, 1500, 2100, and 2400~mg, respectively. The target toxicity and efficacy probabilities were set to $\phi_1=0.30$ and $\de_2=0.5$. The basic parameters of the BOIN and BOIN-ET designs were set as $\phi_2=0.6\phi_1$ for BOIN/iBOIN design and $\phi_2=0.1\phi_1$ for BOIN-ET/iBOIN-ET design, $\phi_3=1.4\phi_1$, and $\de_2=0.6\de_1$ according to the recommended values~\cite{liu2015boin,takeda2018boin}. Assuming an average of six patients per dose level, the initial total sample size was set to 24, and increased to 30 when a new dose level was added. Up to twelve patients were treated at each dose level, and the trial was terminated early if the decision was to remain at the same dose level after treating twelve patients. Dose levels at which the posterior probability that the toxicity probability exceeds $\phi_1$ was greater than 0.95 were eliminated, along with all higher dose levels. Similarly, dose levels were eliminated if the posterior probability that the efficacy probability is below $\de_1$ exceeded 0.99. For the inserted dose, we compared the posterior toxicity probability obtained by borrowing from the skeleton with the empirical toxicity probability estimated from the observed data. If the skeleton-based toxicity probability was substantially smaller than the observed toxicity probability, this indicated that the skeleton underestimated the toxicity at the inserted dose. To avoid inappropriate borrowing in such situations, we introduced a threshold $c$ to quantify the discrepancy between the two estimates. Specifically, when $\ph_T - \ph_{T,\mathrm{borrowing}} > c$, where $\ph_T$ denotes the observed toxicity probability at the inserted dose, we regarded the skeleton-based estimate as inconsistent with the data and did not use it for dose adjustment or MTD selection.
 We examined four values $c \in \{0, 0.1, 0.2, 1\}$. Larger values of $c$ correspond to more permissive borrowing (i.e., skeleton information is discarded less often when it underestimates toxicity), and $c=1$ effectively corresponds to always borrowing. We set the ESS scaling factor to 1. For the online learning approach, we applied the Hedge algorithm. For the BLRM, we assumed the prior distributions
 \begin{align}
 \alpha \sim \mathrm{Normal}(\mu_\alpha,\sigma_\alpha^2),\qquad
\log(\beta) \sim \mathrm{Normal}(\mu_\beta,\sigma_\beta^2).\nonumber
 \end{align}
For the Bayesian-mixture approach combining three candidate skeletons
$r_1,r_2,r_3$, we assigned a Dirichlet prior to the mixing weights
$\boldsymbol{w}=(w_1,w_2,w_3)$: $\boldsymbol{w}\sim\mathrm{Dirichlet}(1,1,1), w_1+w_2+w_3=1$.

For the simulation study, we considered two broad types of cases: fixed cases, which are designed to be intuitive and easy to interpret, and random scenarios, which are included to mitigate the inherent arbitrariness of fixed case specifications.

\subsubsection{Fixed case}
To evaluate the operating characteristics of the proposed methods, we generated data under the three fixed toxicity and three fixed efficacy scenarios of true DLT and efficacy probabilities shown in Table~\ref{tab:sim_scenarios}.

\begin{table}[htbp]
  \centering
  \caption{True DLT and efficacy probabilities under the three simulation scenarios}
  \label{tab:sim_scenarios}
  \begin{tabular}{cccccc}
    \hline
    Scenario & $d_1$ (300 mg) & $d_2$ (900 mg) & $d_3$ (1500 mg) & $d^\ast$ (2100 mg) & $d_4$ (2400 mg) \\
    \hline
    \multicolumn{6}{l}{\textbf{Toxicity}} \\
    T1 & 0.05 & 0.10 & 0.15 & 0.20 & 0.60 \\
    T2 & 0.05 & 0.10 & 0.15 & 0.30 & 0.60 \\
    T3 & 0.05 & 0.10 & 0.15 & 0.50 & 0.60 \\
    \hline
    \multicolumn{6}{l}{\textbf{Efficacy}} \\
    E1 & 0.05 & 0.10 & 0.15 & 0.20 & 0.60 \\
    E2 & 0.05 & 0.10 & 0.15 & 0.30 & 0.60 \\
    E3 & 0.05 & 0.10 & 0.15 & 0.50 & 0.60 \\
    \hline
  \end{tabular}
\end{table}

In all scenarios, the design incorporated historical partial data that had been already observed at the time of adaptation. Specifically, the number of treated patients is $(n_1, n_2, n_3, n^\ast, n_4) = (3, 3, 6, 0, 6)$, the number of observed DLTs is $(y_{T1}, y_{T2}, y_{T3}, y_{T}^\ast, y_{T4}) = (0, 0, 1, 0, 3)$, and the number of observed efficacy is $(y_{E1}, y_{E2}, y_{E3}, y_{E}^\ast, y_{E4}) = (0, 0, 0, 0, 3)$. We assumed that a new intermediate dose, $d^\ast = 2100$~mg, was inserted during the trial, and all simulations were initiated by allocating the next cohort of patients to this newly inserted dose level.

\subsubsection{Random case}
To eliminate arbitrariness in the simulation design, we also conduct random scenario simulations. The construction of these random scenarios follows the simulation settings proposed by Liu and Yuan~\cite{liu2015boin}. The simulation setup is as follows. Only one dose level is added during the trial, and the total number of dose levels after addition is five. When specifying the true toxicity probabilities, we consider the structure consisting of five dose levels including the added one.
\begin{enumerate}
\item One dose level is randomly selected from levels 2 to 4 with equal probability and denoted by $j$. 

\item The true toxicity probabilities at dose levels $j-1$ and $j+1$ are specified as follows:
\begin{align}
p_{j-1}&=\Phi[\ep_j-\{\ep_j-z(2\phi-\Phi(\ep_j))\}I(\ep_j>z(\phi))-\ep^2_{j-1}],\nonumber\\
p_{j+1}&=\Phi[\ep_j+\{z(2\phi-\Phi(\ep_j))-\ep_j\}I(\ep_j<z(\phi))+\ep^2_{j+1}],\nonumber
\end{align}
where $\ep_j\sim\Nc(z(\phi),\si^2_0)$, $z(\cdot)$ denotes the inverse standard normal CDF,
and $\ep_{j-1}\sim\Nc(\mu_1,\si^2_1)$, $\ep_{j+1}\sim\Nc(\mu_2,\si^2_2)$. For the remaining dose levels, $p_{j-k}=\Phi\{z(p_{j-k+1})-\ep^2_{j-k}\},\quad p_{j+l}=\Phi\{z(p_{j+l-1})+\ep^2_{j+l}\},$ where $k$ and $l$ are natural numbers.

\item Dose assignment is performed using the four dose levels removing $j$. If the trial completes without any dose de-escalation during dose assignment, the corresponding simulation replicate is excluded from the analysis. If de-escalation occurs at a dose level $g$ (with $g \ge 2$), a new dose is inserted between levels $g$ and $g-1$. The true toxicity probability of the inserted dose $\ep^*$ is
$\Phi(\ep^*)$, where $\ep^*\sim\Nc((p_{g}-p_{g-1})/2,{\si^*}^2)$.  

\end{enumerate}

We assumed that $\mu_1=\phi-0.5$, $\mu_2=\phi+0.5$, $\si_0=\si^*=0.05$, $\si_1=\si_2=0.5$.

For each simulated trial, we generated a vector of true efficacy probabilities
$\q=(q_1,\ldots,q_J)$ under simple shape constraints, without assuming any
parametric dose--response model~\cite{sun2024statistical}.
Let $q_E$ denote the minimum clinically meaningful efficacy probability and let
$q_{\max}$ be an upper bound.
A dose index $\dt \in \{1,\ldots,J\}$ was first sampled uniformly at random,
and the efficacy probability at that dose was drawn as
\[
q_{\dt} \sim \mathrm{Unif}(\de_1, q_{\max}).
\]
Conditional on $q_{\dt}$, the remaining $q_d$ values were generated using
independent uniform draws constrained to satisfy the prespecified efficacy
shape. We considered two shapes.

\paragraph{Monotone increasing efficacy.}
To construct a nondecreasing efficacy profile, doses below $d^\ast$ were filled
in by drawing successively smaller values:
\[
q_{\dt-1} \sim \mathrm{Unif}(0, q_{\dt}),\quad
q_{\dt-2} \sim \mathrm{Unif}(0, q_{\dt-1}),\ \ldots
\]
Doses above $d^\ast$ were then generated to be successively larger, truncated by
$q_{\max}$:
\[
q_{\dt+1} \sim \mathrm{Unif}(q_{\dt}, q_{\max}),\quad
q_{\dt+2} \sim \mathrm{Unif}(q_{\dt+1}, q_{\max}),\ \ldots
\]
This procedure produces trial-to-trial variability while enforcing the
monotonicity constraint.

\paragraph{Unimodal efficacy.}
To obtain a single-peaked efficacy profile, we additionally sampled a peak dose
index $d^{\mathrm{peak}} \in \{1,\ldots,J\}$ uniformly at random.
If $d^{\mathrm{peak}} \neq d^\ast$, we generated the peak efficacy level from
\[
q_{d^{\mathrm{peak}}} \sim \mathrm{Unif}(q_{\dt}, q_{\max}),
\]
whereas if $d^{\mathrm{peak}} = \dt$ the reference dose also served as the
peak.
Efficacy probabilities were then generated outward from the peak so that they
decrease away from $d^{\mathrm{peak}}$:
\[
q_{d^{\mathrm{peak}}-1} \sim \mathrm{Unif}(0, q_{d^{\mathrm{peak}}}),\quad
q_{d^{\mathrm{peak}}-2} \sim \mathrm{Unif}(0, q_{d^{\mathrm{peak}}-1}),\ \ldots
\]
and similarly on the right side,
\[
q_{d^{\mathrm{peak}}+1} \sim \mathrm{Unif}(0, q_{d^{\mathrm{peak}}}),\quad
q_{d^{\mathrm{peak}}+2} \sim \mathrm{Unif}(0, q_{d^{\mathrm{peak}}+1}),\ \ldots
\]
When $\dt$ and $d^{\mathrm{peak}}$ differ, the above construction is applied
after ensuring that the segment between $d^\ast$ and $d^{\mathrm{peak}}$
remains within $[q_{\dt}, q_{d^{\mathrm{peak}}}]$.
Overall, the resulting $\q$ satisfies a unimodal constraint with random
spacing between adjacent efficacy levels.

In the random scenario simulations, we retain the same data-generating mechanism for toxicity and efficacy as described above. Dose insertion is triggered based on the observed efficacy outcomes: specifically, an intermediate dose is added only when the observed efficacy probability at the lower dose level is below the efficacy threshold $\de_1$
, while the observed efficacy probability at the higher dose level exceeds $\de_1$. This criterion reflects the practical motivation of the proposed method, namely, situations in which efficacy signals are observed at higher dose levels but not at lower ones, prompting the consideration of an intermediate dose.

\subsection{Simulation results}
For the Fixed cases, we report the percentage of correct MTD selection. For designs related to BOIN-ET, we additionally report the percentage of correct OBD selection. We also summarize the selection percentages of each dose as the MTD, the average number of patients treated at each dose level, and, for designs related to BOIN-ET, the selection percentage of each dose as the OBD. In the Supplemental Material, we report the percentage of trials in which a dose higher than the true MTD was selected as the MTD.

For the random scenarios, we similarly report the percentages of correct MTD and OBD selection. The Supplemental Material further includes the percentage of trials in which a dose higher than the true MTD was selected as the MTD.

\subsubsection{Results for fixed case}
\begin{figure}[tbp]\centering
\includegraphics[width=\linewidth]{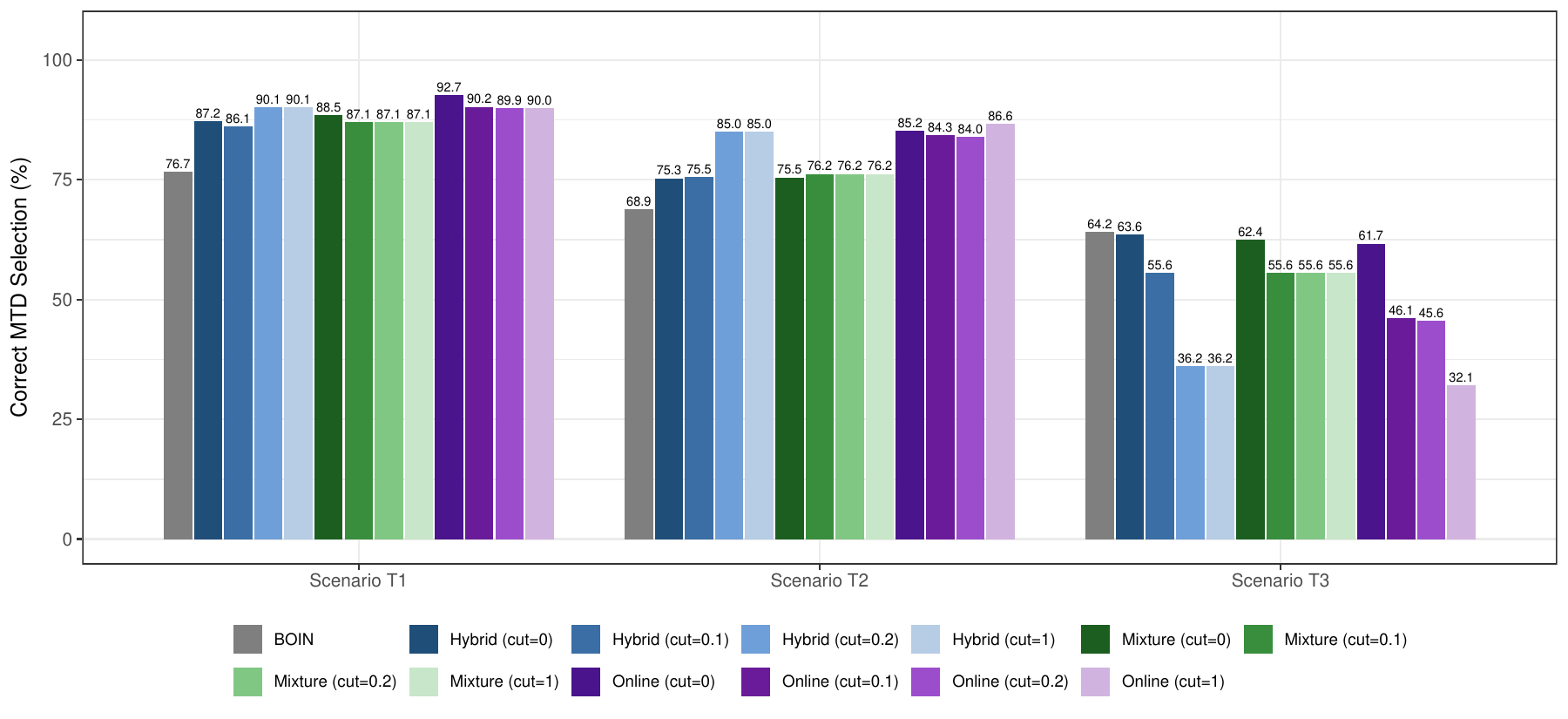}
\caption{Percentage of correct MTD selection for BOIN-related designs in fixed case}
\label{fig:bar_Correct_MTD_Selection_Pct_boundMTDFALSE}
\end{figure}

The simulation results for the BOIN-related designs are presented in Figure \ref{fig:bar_Correct_MTD_Selection_Pct_boundMTDFALSE}. Across all settings in Scenarios 1 and 2, the proposed method achieved a higher percentage of correct MTD selection.
When the inserted dose added during the trial was overly toxic, the proposed method with the threshold $c=0$ deomnstrayed a higher probability of correct MTD selection comparable to the conventional BOIN design.

\begin{figure}[tbp]\centering
\includegraphics[width=\linewidth]{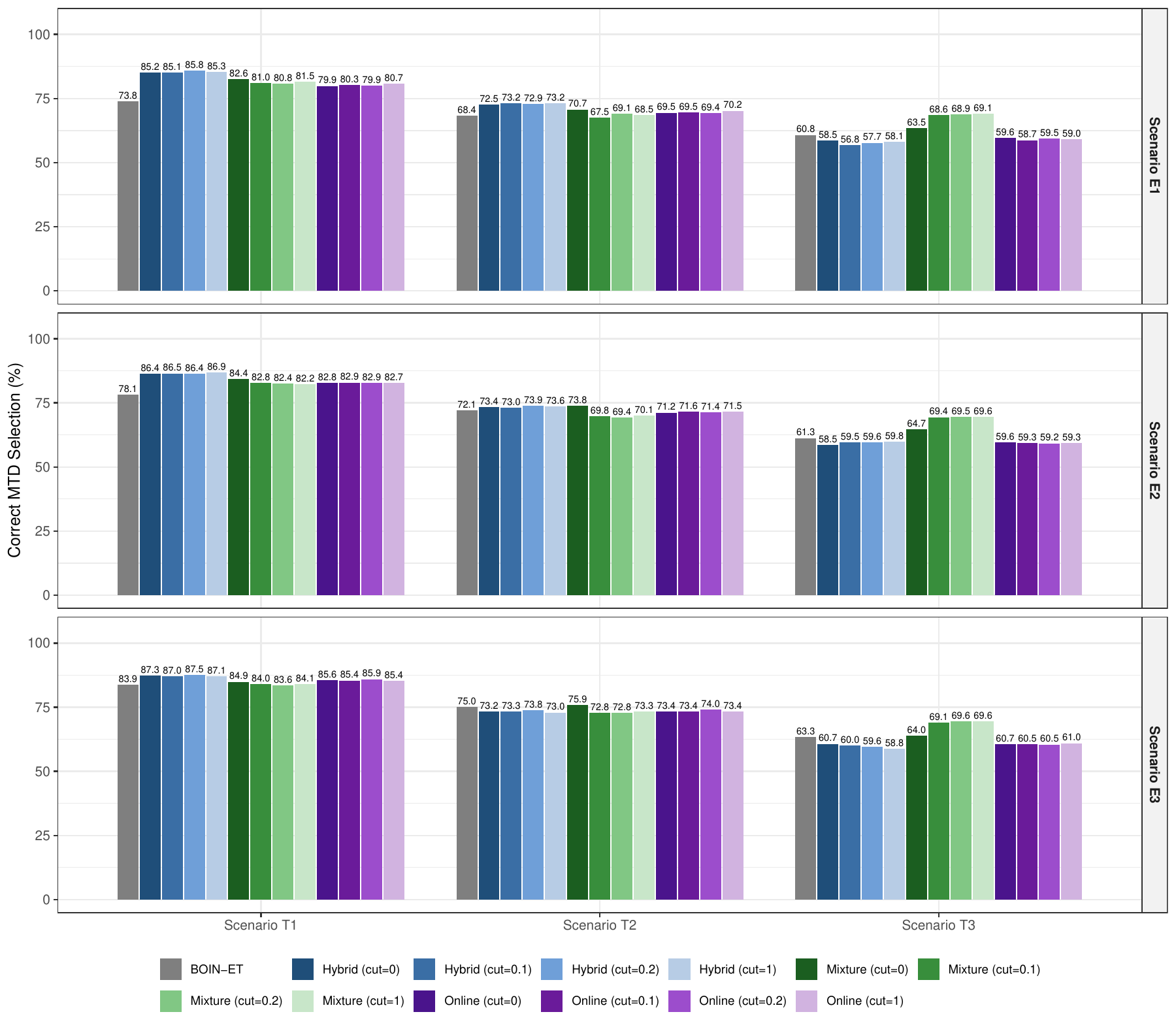}
\caption{Percentage of correct MTD selection for BOIN-ET-related designs in fixed case}
\label{fig:bar_correct_MTD_selection_boundMTDFALSE}
\end{figure}

\begin{figure}[tbp]\centering
\includegraphics[width=\linewidth]{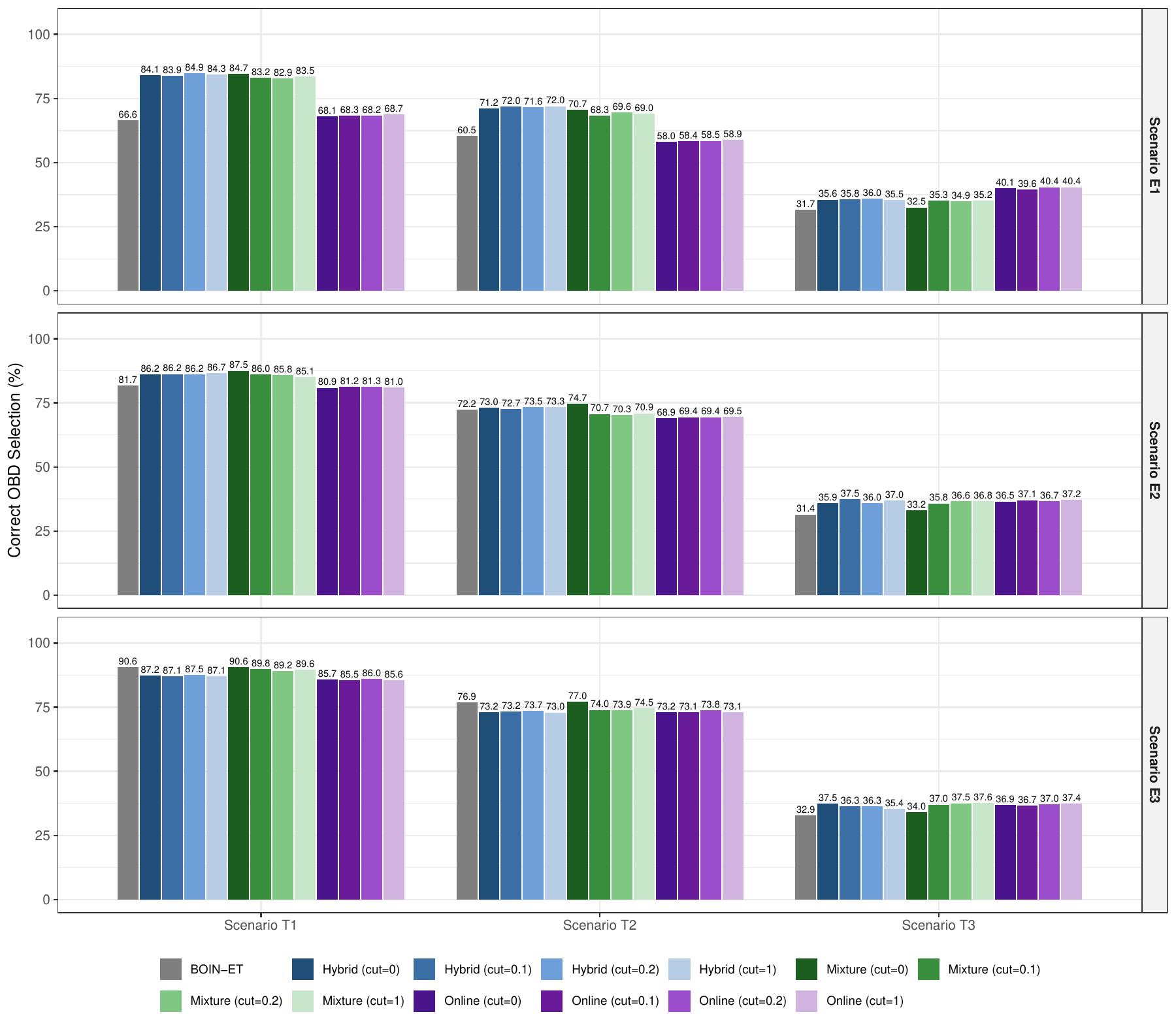}
\caption{Percentage of correct OBD selection for BOIN-ET-related designs in fixed case}
\label{fig:bar_correct_OBD_selection_boundMTDFALSE}
\end{figure}

Next, the simulation results for the BOIN-ET-related designs are presented in Figure \ref{fig:bar_correct_MTD_selection_boundMTDFALSE}. Across all settings in Scenarios 1, the proposed method achieved a higher probability of correct MTD selection. In Scenario 2, the proposed methods demonstrated  a probability of correct MTD selection comparable to that of the conventional BOIN design.
In Scenario 3, where the added intermediate dose was overly toxic, the proposed method outperformed the conventional design for the Mixture approach particularly when cut = 0.1, 0.2, and 1. Figure \ref{fig:bar_correct_OBD_selection_boundMTDFALSE} shows that the percentage of correct OBD selection followed similar patterns to those observed for the MTD.


\subsubsection{Results for random case}
\begin{figure}[tbp]\centering
\includegraphics[width=\linewidth]{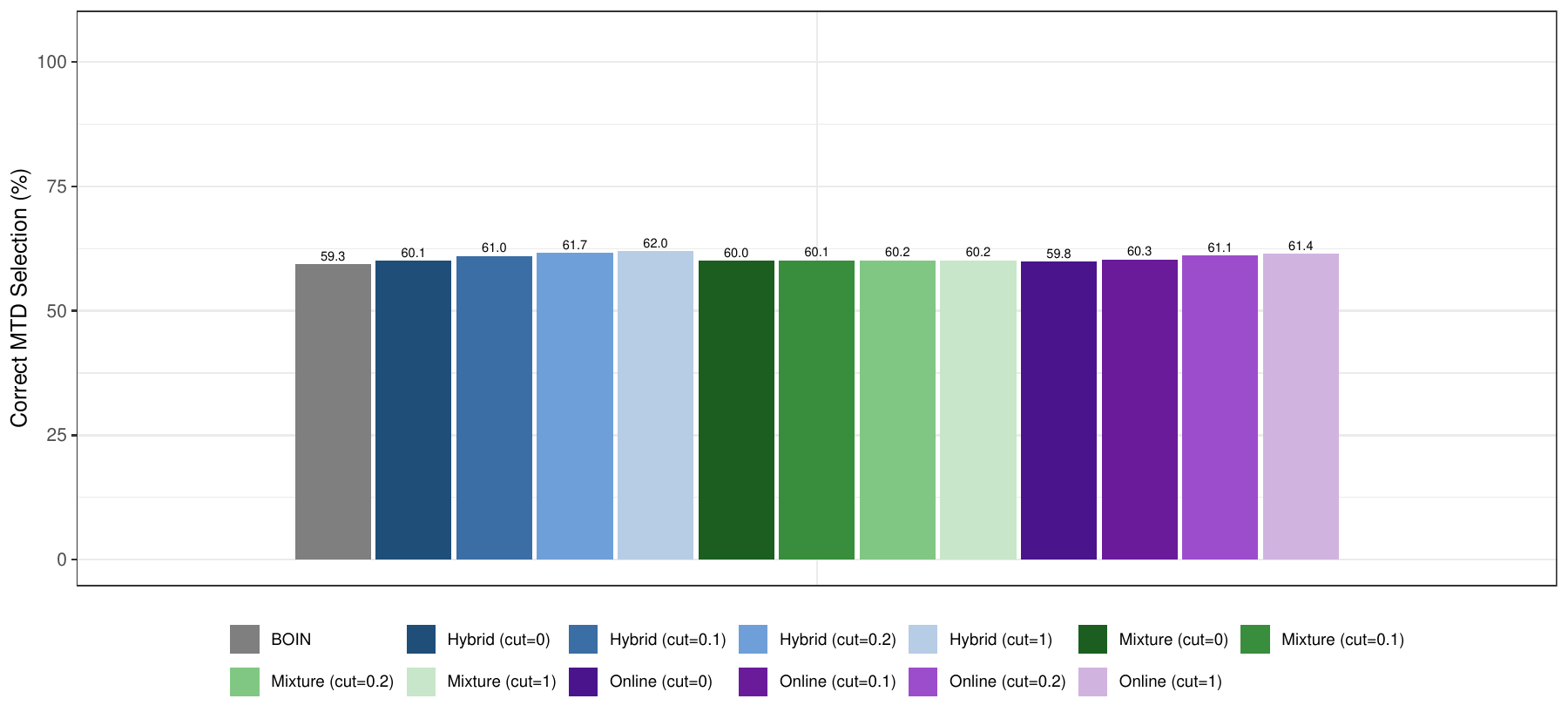}
\caption{Percentage of correct MTD selection for BOIN-related designs in random case}
\label{fig:bar_P_correct_MTD_random_boundMTDFALSE}
\end{figure}

Figure \ref{fig:bar_P_correct_MTD_random_boundMTDFALSE} presents the simulation results for the percentage of correct MTD selection under the random scenarios for the proposed methods related to the BOIN design. Although the proposed method did not perform worse than the original BOIN design, no substantial improvement was observed.

\begin{figure}[tbp]\centering
\includegraphics[width=\linewidth]{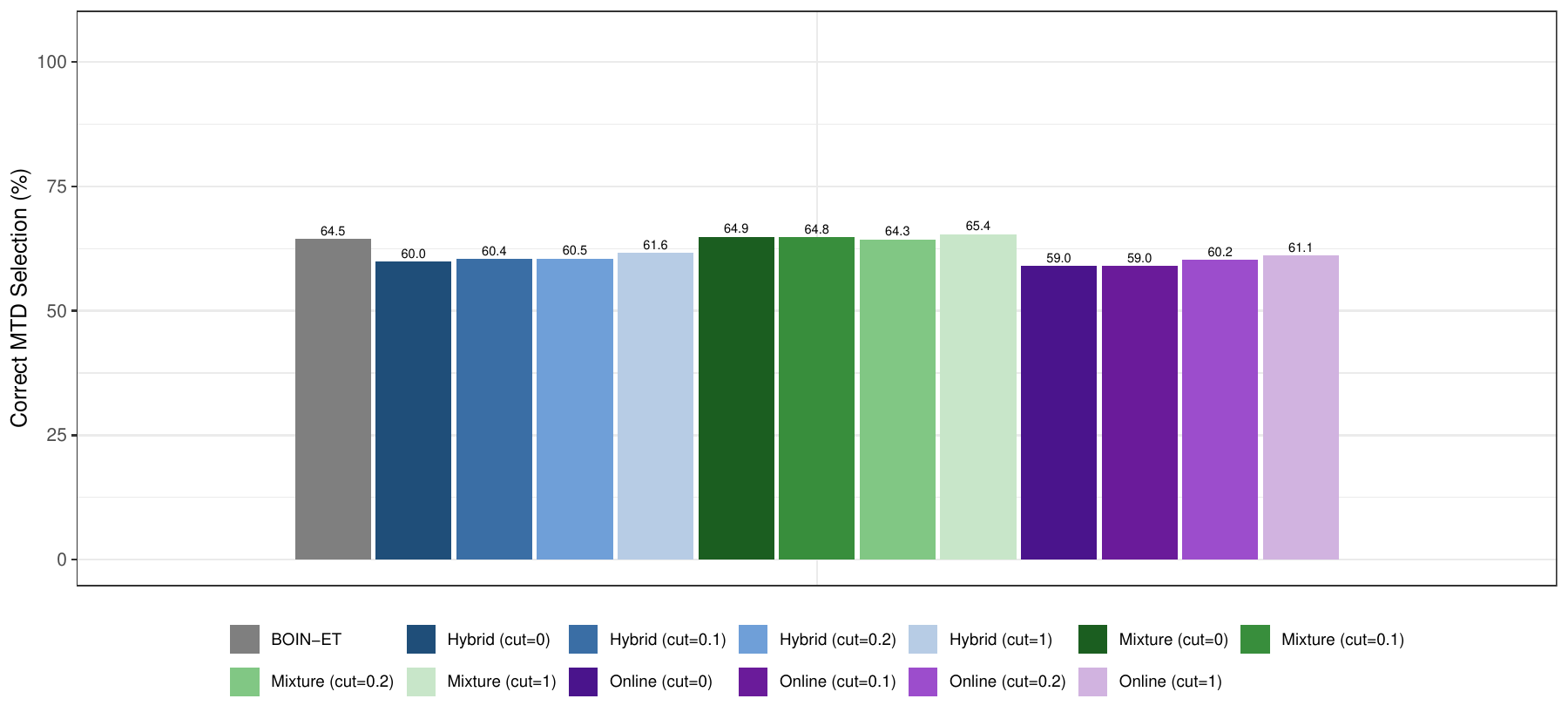}
\caption{Percentage of correct MTD selection for BOIN-ET-related designs under monotonic increasing efficacy curve in random case}
\label{fig:bar_P_correct_MTD_increasing_boundMTD_FALSE}
\end{figure}

\begin{figure}[tbp]\centering
\includegraphics[width=\linewidth]{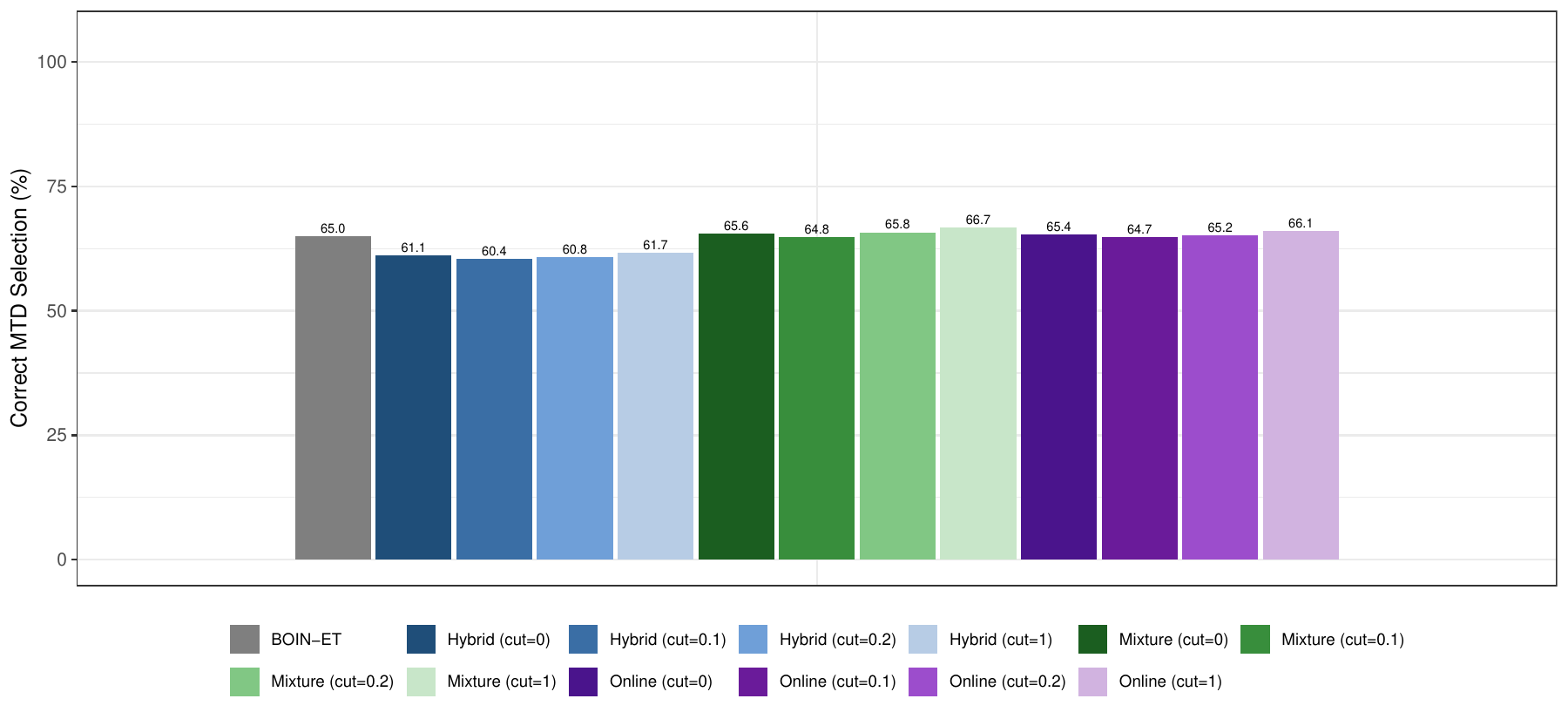}
\caption{Percentage of correct MTD selection for BOIN-ET-related designs under unimodal efficacy curve in random case}
\label{fig:bar_P_correct_MTD_unimodal_boundMTD_FALSE}
\end{figure}

\begin{figure}[tbp]\centering
\includegraphics[width=\linewidth]{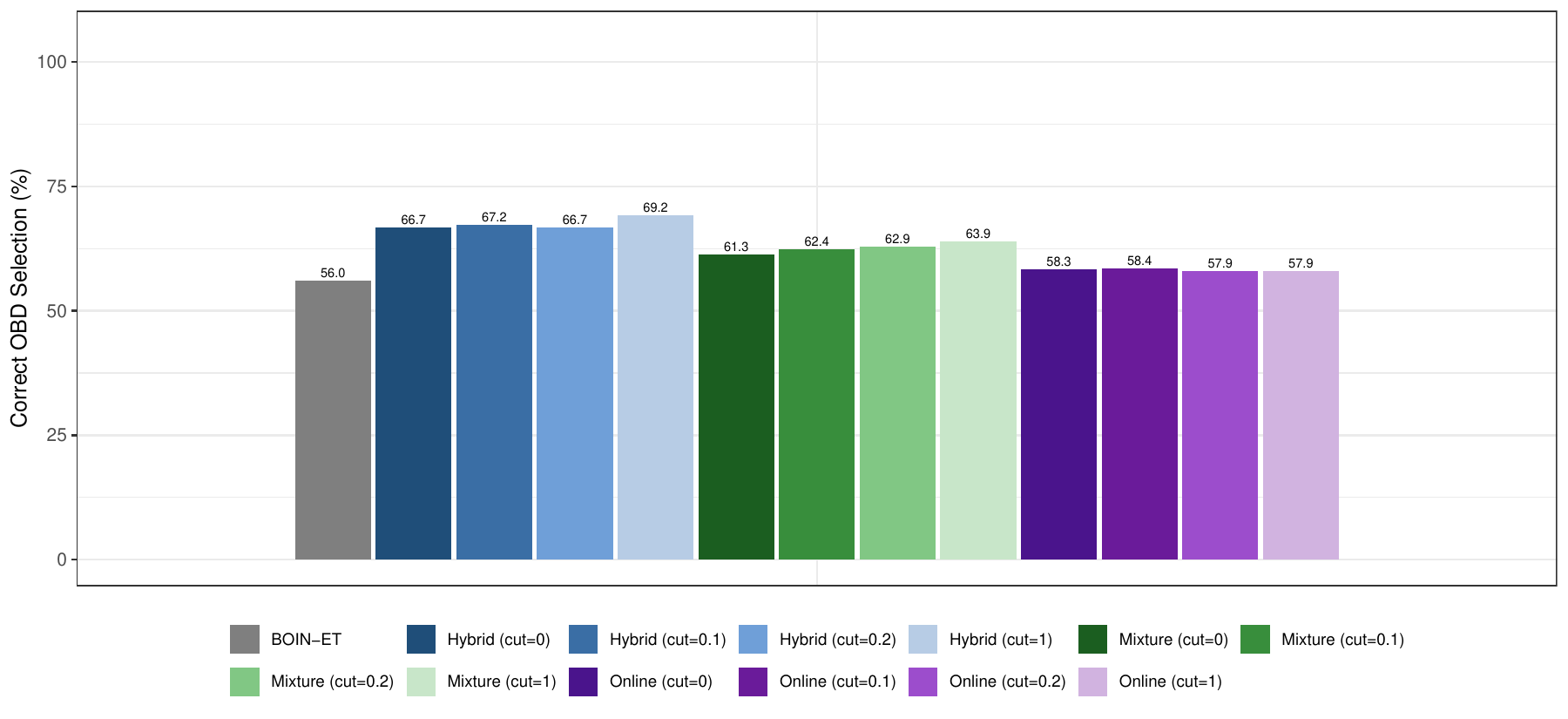}
\caption{Percentage of correct OBD selection for BOIN-ET-related designs under monotonic increasing efficacy curve in random case}
\label{fig:bar_P_correct_OBD_increasing_boundMTD_FALSE}
\end{figure}

\begin{figure}[tbp]\centering
\includegraphics[width=\linewidth]{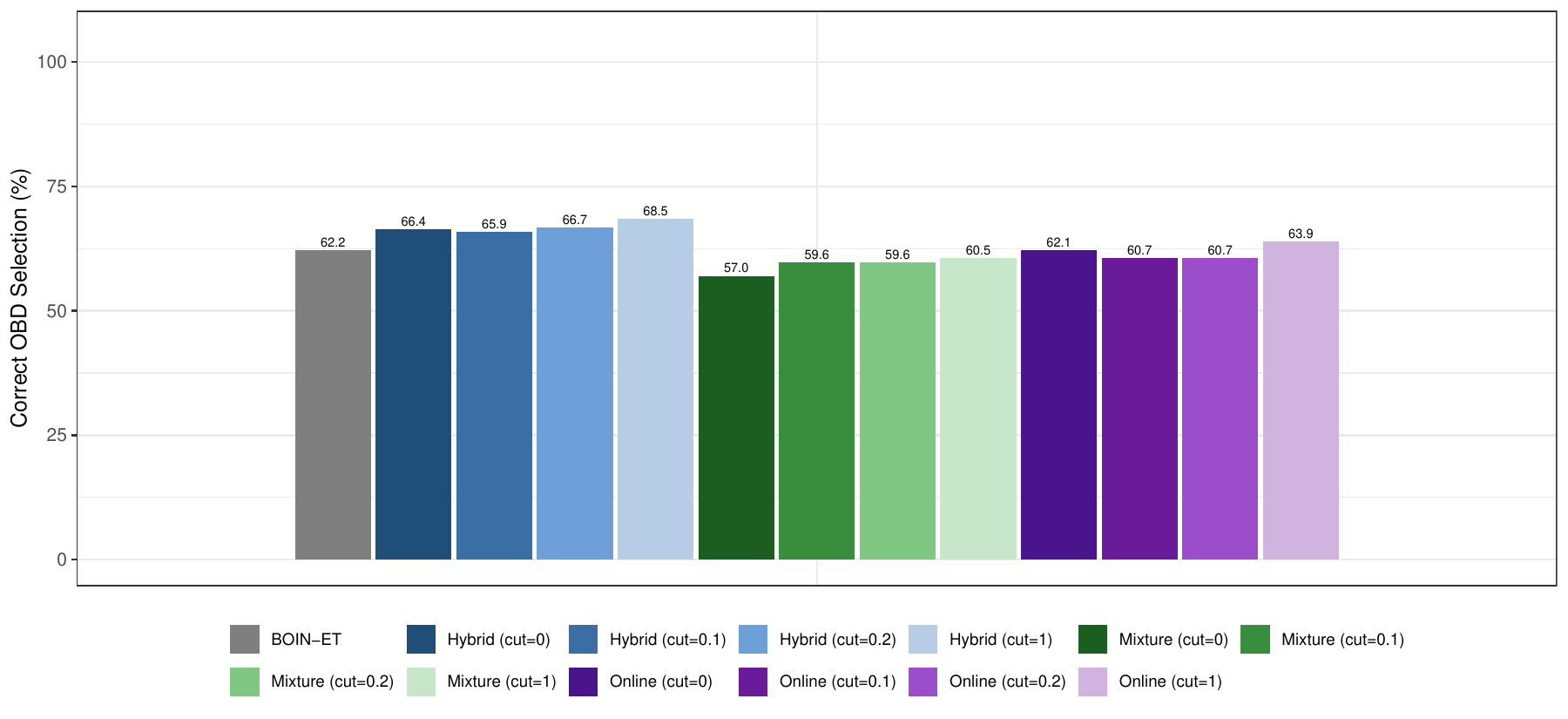}
\caption{Percentage of correct OBD selection for BOIN-ET-related designs under unimodal efficacy curve in random case}
\label{fig:bar_P_correct_OBD_unimodal_boundMTD_FALSE}
\end{figure}

Next, Figures \ref{fig:bar_P_correct_MTD_increasing_boundMTD_FALSE} and \ref{fig:bar_P_correct_MTD_unimodal_boundMTD_FALSE} presents the results for the percentage of correct MTD selection, and Figures \ref{fig:bar_P_correct_OBD_increasing_boundMTD_FALSE} and \ref{fig:bar_P_correct_OBD_unimodal_boundMTD_FALSE} presents the results for the percentage of correct OBD selection, for the BOIN-ET–related designs. Under any efficacy profile assumption, it demonstrated MTD and OBD selection rates almost equivalent to the BOIN-ET design.
In addition, the Supplemental Material reports the proportion of trials judged to be overly toxic for all designs under the random scenarios.

\section{Discussion}
We proposed a hybrid design that combines the BOIN/BOIN-ET designs with their informative extensions, iBOIN/iBOIN-ET, to accommodate mid-trial dose insertion in Phase I oncology trials using the data already observed at the time of insertion. To our knowledge, no previous work has proposed a hybrid strategy that integrates a non-informative model-assisted design with an informative-prior counterpart in this setting, and this paper is the first to apply such a framework.

A key feature of our approach is that, for a newly inserted dose, we leverage information from other dose levels with accumulated data to construct the skeleton and to quantify the strength of borrowing via the ESS. We proposed three approaches to determine the skeleton for the inserted dose: (i) a method based on the Bayesian logistic regression model (BLRM), (ii) a mixture approach that combines the BLRM-based estimate with the empirically observed toxicity/efficacy probabilities at the adjacent dose levels, and (iii) an online-learning approach that adaptively updates the skeleton by incorporating post-insertion data. Because the main contribution of our method lies in estimating the skeleton and ESS for the inserted dose, the framework is not limited to BOIN/BOIN-ET. For example, analogous hybrid designs can be readily developed for other model-assisted designs, such as modified toxicity probability interval (mTPI)~\cite{ji2010modified}/imTPI~\cite{zhou2021incorporating} and Keyboard~\cite{yan2017keyboard}/iKeyboard~\cite{zhou2021incorporating}, by applying the same skeleton/ESS construction step to the newly added dose while retaining the original non-informative rules for the pre-specified doses.

Regarding the simulation results, for the BOIN/iBOIN hybrid design, we found that when the true toxicity probability of the inserted intermediate dose was at or below the TTL, the proposed method achieved a higher probability of correct MTD selection. In contrast, when $c$ was close to $1$, the conventional BOIN design tended to yield a higher probability of correct MTD selection when the inserted dose was overly toxic (i.e., its true toxicity probability exceeded the TTL). Notably, when $c=0$, the fixed-scenario results showed that the proposed method performed comparably to the BOIN design under highly toxic settings, while outperforming BOIN when the true toxicity probability of the intermediate dose was at or below the TTL; a similar tendency was also observed for the online-weighting approach. Meanwhile, the Bayesian-mixture approach was found to improve MTD selection performance while controlling excessive toxicity regardless of the value of $c$. Under the random scenarios, we found that the proposed methods did not perform worse than the original BOIN design, although the improvement in correct MTD selection was not substantial. For BOIN-ET/iBOIN-ET, in the fixed scenarios, we observed a slight performance deterioration in some settings when the inserted intermediate dose had a true toxicity probability at or below the TTL and high efficacy; however, in the other scenarios, the proposed methods demonstrated superior performance. For the online-weighting approach, the performance was not stable, which we attribute to instability in updating efficacy, whereas the hybrid and Bayesian-mixture approaches exhibited stable and favorable trends. In the random scenarios, under a monotonically increasing efficacy curve, the hybrid and Bayesian-mixture approaches performed well, and under a unimodal efficacy curve, the hybrid approach showed good performance. Finally, the simulations clearly indicated that excessive toxicity was well controlled.

In conclusion, for the BOIN/iBOIN hybrid design, setting the borrowing threshold to $c=0$ provided a favorable trade-off: it improved MTD selection when the inserted dose was at or below the TTL while maintaining performance comparable to BOIN in highly toxic settings. In addition, when using the Bayesian-mixture approach, we found that excessive toxicity can be largely avoided even without imposing a borrowing threshold. Moreover, for the BOIN-ET design, we confirmed that the proposed method increases the probability of correct OBD selection while preventing excessive toxicity.

\newpage
\bibliography{main.bib}
\bibliographystyle{unsrt}

\newpage
%\begin{appendix}
%\section{App}
%\label{appa}
%In this section, 
%\end{appendix}

\end{document}